\documentclass[aps,prx,twocolumn,showpacs,superscriptaddress]{revtex4}
\usepackage{amsmath}
\usepackage{amssymb}
\usepackage{epsfig}
\usepackage{color}
\usepackage{graphicx,amsmath}

\begin{document}

\title{Flat bands and enigma of metamagnetic quantum critical regime in
$\rm \bf Sr_3Ru_2O_7$}
\author{V. R. Shaginyan}\email{vrshag@thd.pnpi.spb.ru}
\affiliation{Petersburg Nuclear Physics Institute, Gatchina,
188300, Russia}\affiliation{Clark Atlanta University, Atlanta, GA
30314, USA} \author{A. Z. Msezane}\affiliation{Clark Atlanta
University, Atlanta, GA 30314, USA}\author{K. G.
Popov}\affiliation{Komi Science Center, Ural Division, RAS,
Syktyvkar, 167982, Russia}
\author{J.~W.~Clark}
\affiliation{McDonnell Center for the Space Sciences \& Department
of Physics, Washington University, St.~Louis, MO 63130, USA}
\author{M.~V.~Zverev}
\affiliation{Russian Research Centre Kurchatov Institute, Moscow,
123182, Russia} \affiliation{Moscow Institute of Physics and
Technology, Moscow, 123098, Russia}
\author{V. A. Khodel} \affiliation{Russian Research Centre Kurchatov Institute,
Moscow, 123182, Russia} \affiliation{McDonnell Center for the Space
Sciences \& Department of Physics, Washington University,
St.~Louis, MO 63130, USA}

\begin{abstract}
Understanding the nature of field-tuned metamagnetic quantum
criticality in the ruthenate $\rm Sr_3Ru_2O_7$ has presented a
significant challenge within condensed matter physics. It is known
from experiments that the entropy within the ordered phase forms a
peak, and is unexpectedly higher than that outside, while the
magnetoresistivity experiences steep jumps near the ordered phase.
We find a challenging connection between $\rm Sr_3Ru_2O_7$ and
heavy-fermion metals expressing universal physics that transcends
microscopic details. Our construction of the $T-B$ phase diagram of
$\rm Sr_3Ru_2O_7$ permits us to explain main features of the
experimental one, and unambiguously implies an interpretation of
its extraordinary low-temperature thermodynamic in terms of fermion
condensation quantum phase transition leading to the formation of a
flat band at the restricted range of magnetic fields $B$. We show
that it is the flat band that generates both the entropy peak and
the resistivity jumps at the QCPs.\end{abstract}


\maketitle

{Corresponding author: V. R. Shaginyan at: Petersburg Nuclear
Physics Institute, Gatchina, 188300, Russia; Phones: (office)
7-813-714-6096, (fax) 7-813-713-1963; E-mail address: vrshag@thd.pnpi.spb.ru}\\

\section{Introduction}

Discoveries of surprising and exotic phenomena in strongly
correlated metals provide unique opportunities for expanding our
understanding of quantum critical physics. A case in point is the
quantum critical metal $\rm Sr_3Ru_2O_7$, a member of the
Ruddlesden-Popper series of layered perovskite ruthenates
consisting of $\rm RuO_2$ $ab$ planes forming bilayers which are
piled along the crystalline $c$ axis, perpendicular to the $ab$
axis, and weakly coupled to one another. In spite of numerous
experimental and theoretical investigations
\cite{sc1,per,sc4,sc7,doping,sc9,ragt,natp,pnas,millis,sigr,nj_mac},
explanation of the puzzling low-temperature behavior of this
material in external magnetic fields $B$ remains an open problem in
condensed matter theory, for recent reviews see e.g.
\cite{frad,physc}. The observations indicate that the physics
underlying this behavior of $\rm Sr_3Ru_2O_7$, which resembles that
of some heavy-fermion (HF) metals, is not subsumed in the
spin-Kondo picture \cite{ragt,natp,pnas}. In crystals of high
quality with residual resistivity $\rho_{\rm res}\sim 0.4\,{\rm
\mu\Omega\, cm}$ measured at zero $B$-field and mean free path
approximately 3000 {\AA}, one observes a metamagnetic transition
featuring a sudden and sharp rise in the magnetization $\mathbf{M}$
with a modest increase in the applied field \cite{frad,physc},
accompanied by a bifurcation of the metamagnetic phase boundary. At
low temperatures the bifurcation splits into two first-order
metamagnetic transitions at critical magnetic field values
$B_{c1}\simeq7.8$ T and $B_{c2}\simeq 8.1$ T
\cite{per,sc4,frad,physc}. Conventionally, a phase that emerges at
fields $B_{c1}<B<B_{c2}$ and temperatures $T\leq T_c\simeq1.2$ K is
identified as a nematic one \cite{sc4,sc7,sc9}. This phase breaks
the discrete square lattice rotational symmetry, as witnessed by a
large magnetoresistive anisotropy in the $ab$ plane as the magnetic
field $B$ is rotated away from the $c$ axis toward the $ab$ plane
\cite{sc4,sc7}. The anisotropy vanishes as soon as the $B$-field is
directed along the $c$ axis (here we consider only this $B
\parallel c$ case). The two first-order transitions persist, but
convert into two second second-order phase transitions with rising
temperature as illustrated in Fig.~\ref{fig1} {\bf A}.

In our letter, we reveal a challenging connection between $\rm
Sr_3Ru_2O_7$ and heavy-fermion metals expressing universal physics
that transcends microscopic details. Our construction of the $T-B$
phase diagram of $\rm Sr_3Ru_2O_7$ permits us to explain main
features of the experimental one. The obtained agreement with the
experimental phase diagram is robust and does not depend on the
nature of the ordered phase for our analysis is based on the
thermodynamic consideration. We show that it is the flat band that
generates both the entropy peak and the resistivity jumps at the
QCPs. We reveal the nature of quantum critical points (QCPs) as
associated with the fermion condensation quantum phase transition
(FCQPT) \cite{shagrep}, and demonstrate that these steep jumps are
represented by jumps of irregular residual resistivity $\rho_0^c$
due to the presence of flat band, generated by FCQPT and leading to
fermion condensation (FC), rather than by the ordered phase itself.
The same QCPs make the entropy within the ordered phase form a peak
and produce a scaling behavior of the thermodynamic functions
strongly resembling that of HF metals.

\section{Magnetoresistivity}

\begin{figure}[!ht]
\begin{center}
\includegraphics [width=0.40\textwidth]{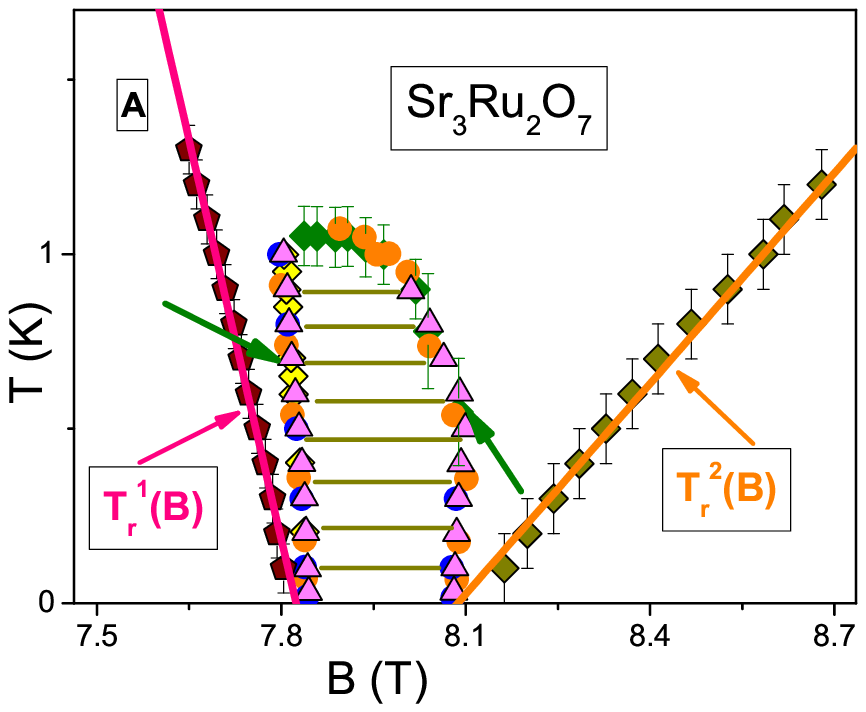}
\includegraphics [width=0.40\textwidth]{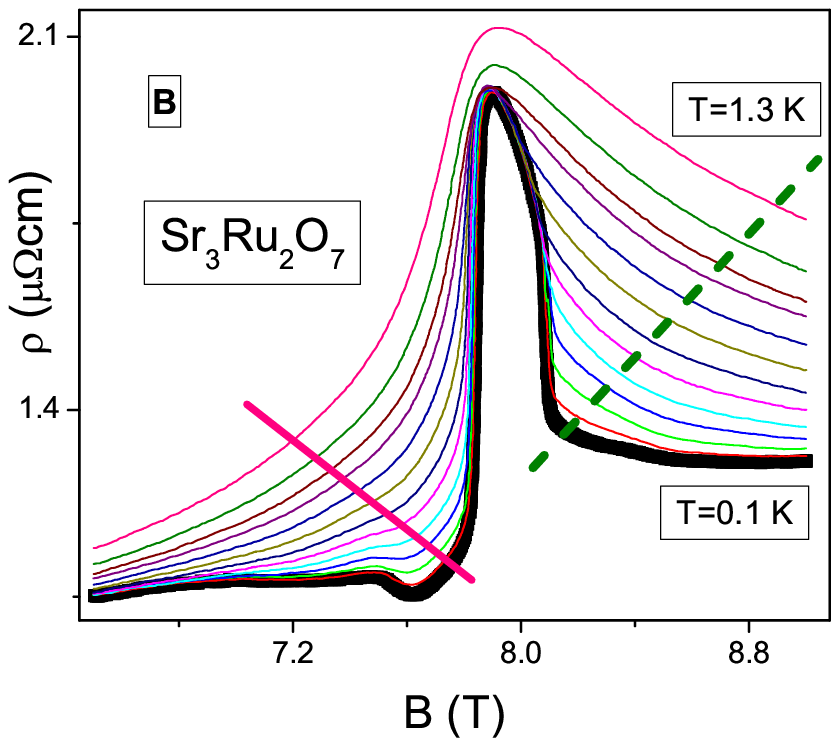}
\end{center}
\caption{(Color online). Panel {\bf A}.  Experimental phase diagram
of $\rm Sr_3Ru_2O_7$ in the $T-B$ plane with magnetic field B as
the control parameter. Geometric symbols representing data points
derived from measurements of susceptibility, magnetostriction,
thermal expansion and transport measurements, surround an area
cross-hatched with horizontal lines representing transitions
between equilibrium thermodynamic phases associated with a nematic
ordered phase \cite{sc4,sc7}. This phase is entered by first-order
phase transitions at low temperatures (indicated by arrows) and by
second-order phase transitions at high temperatures. The solid
lines labeled with $T^1_{r}(B)$ and $T^2_{r}(B)$ on the low- and
the high-field sides of the plot, respectively, sketch the
temperature-field dependence $T^1_{r}(B)$ and $T^2_{r}(B)$ of the
crossings of the resistivity $\rho(B,T)$ drawn in panel {\bf B}
with the two straight lines shown in the same panel. Panel {\bf B}.
The measured resistivity \cite{sc4} $\rho(B)$ of $\rm Sr_3Ru_2O_7$
at a series of temperatures between 0.1 and 1.3 K, in steps of 100
mK. The two lines cross the curves $\rho(B)$ at points symbolized
by pentagons and diamonds in panel {\bf A} and define the crossings
$T^1_{r}(B)$ and $T^2_{r}(B)$, respectively.}\label{fig1}
\end{figure}

The magnetoresistivity $\rho(B,T)$ as a function of field and
temperature is frequently approximated by the formula
\begin{equation}
\rho(B,T)=\rho_{\rm res}+\Delta\rho(B)+AT^n,\label{res}
\end{equation}
where $\Delta\rho$ is the correction to the resistivity produced by
the field $B$ and $A$ is a $T$-independent coefficient. The index
$n$ takes the values 2 and 1, respectively, for Landau Fermi liquid
(LFL) and non-Fermi liquid (NFL) states and values $1\lesssim
n\lesssim 2$ in the crossover region between them. Although both
the resistivity anisotropy and the ordered phase are striking, the
essential physics in $\rm Sr_3Ru_2O_7$ seems to be that of a
normal-phase electron fluid at fields $B<B_{c1}$ and $B>B_{c2}$,
and at $T>T_c$ and $B_{c2}>B>B_{c1}$, i.e., outside of the ordered
phase. Indeed, data on the low-temperature magnetoresistivity
$\rho(B)$ with $B\parallel c$ collected for $\rm Sr_3Ru_2O_7$
\cite{sc4} show that large changes of $\rho(B)$ occur as the
ordered phase is approached. The straight lines indicated by arrows
in Fig.~\ref{fig1} {\bf A}, crossing pentagon data points on the
low-field side and diamonds on the high-field side, represent the
functions $T^1_r(B)$ and $T^2_r(B)$, respectively. In reference to
panel {\bf B} of Fig.~\ref{fig1}, which shows the resistivity
versus field strength at a series of temperatures, the straight
line of panel {\bf A} crossing pentagons [respectively, diamonds]
delineates the intersection with the solid [dashed] line appearing
in panel {\bf B}. It is seen from Fig. \ref{fig1} panel {\bf A},
that $T^1_{r}(B=B_{c1})\simeq0$ and $T^2_{r}(B=B_{c2})\simeq0$.
Thus, these functions show that the low-temperature sides of
$\rho(B)$ at $T\to 0$ approach the steep sidewalls of the
first-order phase transitions depicted in Fig.~\ref{fig1} {\bf A}.
Accordingly $\rho(B)$ possesses two steep sidewalls as the critical
fields $B_{c1}$ and $B_{c2}$ are approached at $T\to0$, a behavior
evident in both panels of Fig.~\ref{fig1}.

The behavior at $B\simeq 7.9$ T and $T>T_c$ is equally striking in
that $\rho$ is precisely linear in T at least over the range
$1.2\leq T\lesssim18$ K\,\, \cite{pnas}.  This fact allows us to
estimate the irregular residual resistivity $\rho_0^c(B)$. To
evaluate $\rho_0^c(B)$, we extrapolate the data on the resistivity
$\rho$\,\, \cite{sc4} to zero temperature as if the ordered phase
were absent in the range $B_{c2}>B>B_{c1}$.
In this way at $B=7.9$ T one finds $\rho_0^c\sim 1.7\,{\rm
\mu\Omega\, cm}$, while the residual resistivity in the presence of
the nematic phase is $\rho_0^{\rm nem}\sim 2.0\,{\rm \mu\Omega\,
cm}$ in this range. On the other hand, at $B\simeq 7.6$ T the
residual resistivity is $\rho_0^c(B)\simeq 1.1\,{\rm \mu\Omega\,
cm}$. Thus, even in the absence of the ordered phase, the field $B$
triggers an upward jump of the resistivity upon approaching
$B_{c1}$ from below.  The resistivity is approximately constant in
the range $B_{c1}<B<B_{c2}$ and undergoes a second jump downward as
$B$ approaches $B_{c2}$.  Such behavior, seen at the lowest
accessible temperatures of 15 mK \cite{gegs} and 70 mK \cite{hall},
at which the term $AT^n$ in Eq. \eqref{res} can be safely omitted,
is consistent with both the jumps at QCPs and the constancy in the
range $B_{c1}<B<B_{c2}$ of the irregular {\it residual} resistivity
$\rho_0^c$. We conclude that it is $\rho_0^c$ that are responsible
for the observed behavior of $\rho$.

Let us assess possible causes for the jumps. When considering
spin-orbit coupling in disordered electron systems where electron
motion is diffusive, the residual resistivity may have both
positive (weak localization) and negative (weak anti-localization)
signs \cite{larkin}. However, since $\rm Sr_3Ru_2O_7$ exhibits
successive upward and downward jumps separated by the narrow range
of magnetic fields $B_{c2}-B_{c1}$, it is unclear how weak
localization at $B_{c1}$ is changed to weak anti-localization at
$B_{c2}$. Moreover, $\rm Sr_3Ru_2O_7$ is one of the purest metals
and the applicable regime of electron motion is ballistic rather
than diffusive. Therefore, both weak and anti-weak localization
scenarios are irrelevant. Accordingly, one expects the
$B$-dependent correction $\Delta\rho$ to the residual resistivity
to be positive and small. As we have seen, this is far from the
case.

Proposals for the origin of the jumps at finite temperatures may
invoke band electrons close to a van Hove singularity (vHs), giving
rise to the ordered phase \cite{per,sc7,sigr,frad,physc,ragt}.
However, such scenarios must be rejected, since the electrons
involved must have very large effective mass $M^*$ and hence
contribute only weakly to transport properties at finite
temperatures and not at all at $T=0$, see e.g. Ref.
\onlinecite{ragt}. Another possible source of the observed jumps
might be resistivity associated with nematic domains that are
thought to exist at $T\leq T_c$ for the fields tuning the system to
a vHs\, \cite{ragt,sc7,merc,geg4}. The jumps would be due to the
extra scattering produced by such domains. This scenario is also
problematic, since the domains would have to be present at least in
the normal phase and at temperatures as high as $T\sim 18$ K, while
the critical temperature for formation of both the nematic phase
and the domains is estimated as $T_c\simeq 1.2$ K
\cite{sc7,merc,geg4}. Failing such conventional explanations, we
are faced with a challenging task that may well have broad
implications for our understanding of unorthodox (notably, NFL)
phenomena condensed-matter systems: how does one unveil the QCPs
that create the quantum critical regime of $\rm Sr_3Ru_2O_7$,
giving rise to the emergence of jumps in the resistivity $\rho_0^c$
and generating the entropy peak?

\section{Fermion condensation}

In order to develop viable explanations of the resistivity jumps
and entropy excess, it is necessary to recall the origin, nature,
and consequences of {\it flattening} of single-particle excitation
spectra $\varepsilon({\bf p})$ (``flat bands'') in strongly
correlated Fermi systems -- also called swelling of the Fermi
surface or FC \cite{khod,khs,volovik,noz,volov} (for recent
reviews, see \cite{shagrep,shag,mig100}). At $T=0$, the ground
state of a system with a flat band is degenerate, and the
occupation numbers $n_0({\bf p})$ of single-particle states
belonging to the flat band are continuous functions of momentum
${\bf p}$, in contrast to discrete standard LFL values 0 and 1.
Such behavior of the occupation numbers leads to a $T$-independent
entropy term
\begin{equation}
S_0=-\sum_{\,{\bf p}} [n_0({\bf p})\ln n_0({\bf p})+(1-n_0({\bf
p}))\ln(1- n_0({\bf p}))] \label{S0}
\end{equation}
that does not contribute to the specific heat $C(T)=TdS/dT$. Unlike
the corresponding LFL entropy, which vanishes linearly as $T\to 0$,
the term $S_0$ produces a $T$-independent thermal expansion
coefficient \cite{shagrep,aplh,plas,prb}. That $T$-independent
behavior is observed in measurements on $\rm CeCoIn_5$
\cite{oes,steg,zaum} and $\rm YbRh_2(Si_{0.95}Ge_{0.05})_2$
\cite{kuch}, while very recent measurements on $\rm Sr_3Ru_2O_7$
indicate the same behavior \cite{gegprl,gegen} and confirm the
existence of flat bands \cite{nj_mac}. In the theory of fermion
condensation, the degeneracy of the NFL ground state is removed at
any finite temperature, since the flat band acquires a small
dispersion \cite{noz}
\begin{equation}
\varepsilon({\bf p})=T\ln \frac{1-n_0({\bf p})}{n_0({\bf p})}
\label{tem}
\end{equation}
proportional to $T$. The occupation numbers $n_0$ of FC remain
unchanged at relatively low temperatures and, accordingly, so does
the entropy $S_0$.

We now introduce these concepts to achieve a coherent picture of
the quantum critical regime underlying the jump phenomena in $\rm
Sr_3Ru_2O_7$.  In constructing a field-induced flat band, we employ
the model \cite{sc7,sc9,ragt,sigr,berr,puet} based on a vHs that
induces a peak in the single-particle density of states (DOS) and
sharp rise of $\mathbf{M}$ as the field sweeps across the
metamagnetic transition. Upon increase of an applied magnetic field
$B$, the vHs is moved through the Fermi energy. At fields in the
range $B_{c1}<B<B_{c2}$ the DOS peak turns out to be at or near the
Fermi energy. A key point in this scenario is that within the range
$B_{c1}<B<B_{c2}$, a relatively weak repulsive interaction (e.g.,
Coulomb) is sufficient to induce FC and formation of a flat band
with the corresponding DOS singularity locked in the Fermi energy
\cite{shagrep,shag,mig100,yudin}. Now, it is seen from Eq.
\eqref{tem} that finite temperatures, while removing the degeneracy
of the FC spectrum, do not change $S_0$, threatening the violation
of the Nernst theorem.  To avoid such an entropic singularity, the
FC state must be altered as $T\to 0$, so that the excess entropy
$S_0$ is shed before zero temperature is reached. This can take
place by means of some phase transition or crossover, whose
explicit consideration is beyond the scope of this paper.
\begin{figure}[!ht]
\begin{center}
\includegraphics [width=0.47\textwidth]{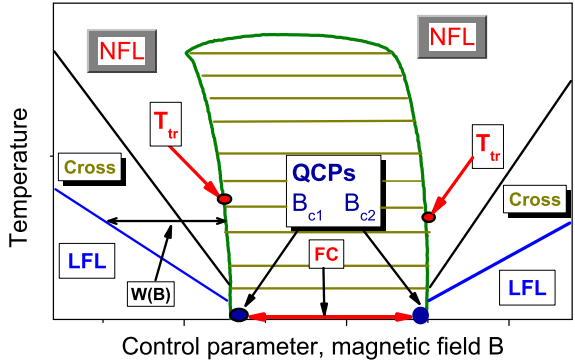}
\end{center}
\caption{(Color online). Schematic phase diagram of the metal $\rm
Sr_3Ru_2O_7$. The QCPs situated at the critical magnetic fields
$B_{c1}$ and $B_{c2}$ are indicated by arrows. The fermion
condensation (FC) or flat band is present between these QCPs as
depicted by the double-headed arrow. The ordered phase bounded by
the curve and demarcated by horizontal lines emerges to eliminate
the entropy excess given by Eq. \eqref{S0}. Two arrows label the
tricritical points $T_{\rm tr}$ at which the lines of second-order
phase transitions change to the first order. The total width of the
NFL state and the crossover leading to the LFL state, $W(B)\propto
T$, is denoted by an arrow. The LFL state occurs at the lowest
temperatures below and above the critical values $B_{c1}$ and
$B_{c2}$ of the tuned $B$-field. Rising temperature ushers in a
broad crossover (labeled Cross) from the LFL to the NFL state.}
\label{fig2}
\end{figure}

\section{Phase diagram}

The schematic $T-B$ phase diagram of $\rm Sr_3Ru_2O_7$ inferred
from the proposed scenario is presented in Fig.~\ref{fig2}.  Its
main feature is the magnetic field-induced quantum critical domain
created by QCPs situated at $B_{c1}$ and $B_{c2}$ and supporting a
FC and associated flat band induced by a vHs (double-headed arrow
between black dots).  The low-field [high-field] QCP on the left
[right] augurs the emergence of the flat band as $B\to B_{c1}$
[$B\to B_{c2}$] from below [above].  In contrast to the typical
phase diagram of a HF metal \cite{shagrep}, the domain occupied by
the ordered phase in Fig.~\ref{fig2} is seen to be approximately
symmetric with respect to the magnetic field
$B_c=(B_{c2}+B_{c1})/2$. The emergent FC and QCPs are considered to
be hidden or concealed in a phase transition, which is driven by
the need for the system to avoid the entropic singularity that
would be produced at $T\to 0$ by the $T$-independent entropy term
$S_0$ of Eq. \eqref{S0}.  The area occupied by this phase
transition is indicated by horizontal lines and restricted by
boundary lines (``sidewalls'' and ``roof''). At the critical
temperature $T_c$ where the new (ordered) phase sets in, the
entropy is a continuous function. Therefore the "roof" of the
domain occupied by the new phase is a line of second-order phase
transitions. As $T$ is lowered, some temperature $T_{\rm tr}$ is
reached at which the entropy of the ordered phase becomes larger
than that of the adjacent disordered phase, due to the remnant
entropy $S_0$ from the highly entropic flat-band state that is
present above $T_c$. Therefore, under the influence of the magnetic
field, the system undergoes a first-order phase transition upon
crossing a sidewall boundary at $T=T_{\rm tr}$, since entropy
cannot be equalized there. It follows, then, that the line of
second-order phase transitions is changed to a line of first-order
transitions at tricritical points indicated by arrows in
Fig.~\ref{fig2}.  It is seen from Fig.~\ref{fig2} that the sidewall
boundary lines are not strictly vertical, due to the stated
behavior of the entropy at the boundary and as a consequence of the
magnetic Clausius-Clapeyron relation \cite{sc9,ragt}. Indeed, in
our case the Clausius-Clapeyron equation reads,
\begin{equation}\label{CCL}
\mu_0\frac{dB_{cn}}{dT_c}=-\frac{\Delta S}{\Delta \mathbf{M}}.
\end{equation}
Here, $\mu_0$ is the permeability constant and $B_{cn}$ stands for
$B_{c1}$ and $B_{c2}$. The beauty of Eq. \eqref{CCL} is that it
defines the slope of the boundary lines shown in Fig. \ref{fig2}
from the first principles of thermodynamics. Since the entropy $S$
within the bounded region is higher than that outside, the slopes
of the phase boundaries point outwards as shown in Fig. \ref{fig2}.
We conclude that the phase diagram \ref{fig2} is in good agreement
with the experimental one shown in Fig. \ref{fig1}, Panel {\bf A}.
We note that the obtained agreement is robust and does not depend
on the nature of the ordered phase for our analysis is based on the
thermodynamic consideration. For example, such a consideration
allows one to establish the $T-B$ phase diagram of the HF metal
$\rm CeCoIn_5$ that resembles that of $\rm Sr_3Ru_2O_7$
\cite{prb,eplce}.

On each flank of the region occupied by the ordered (nematic)
phase, the system crosses over from the LFL state prevailing at the
lowest temperatures to a NFL state under rising temperature. The
total width $W(B)$ of the NFL state and crossover (``Cross'')
region on either flank (denoted with a double arrow in
Fig.~\ref{fig2}), is proportional to $T$\,\, \cite{prb,jetpl}. The
behavior of $W(B)$ inferred from this phase diagram is also
reflected in Fig.~\ref{fig1} {\bf B}, which depicts the dependence
of the function $\rho(B)$ on field strength and temperature.  Since
the width $W(B)$ vanishes when the magnetic field tends to its
critical values, $\rho(B)$ is represented by the two steep
sidewalls seen in panel {\bf B} of Fig.~\ref{fig1} as the critical
field values $B_{c1}$ and $B_{c2}$ are respectively approached from
below and above.

\section{Jumps}

We turn next to calculations of the resistivity $\rho$ in the range
$B_{c1}<B<B_{c2}$, the dispersion of the flat band being governed by
Eq. \eqref{tem}. The electronic liquid of $\rm Sr_3Ru_2O_7$ is
described by several bands occupied by normal quasiparticles that
simultaneously intersect the Fermi surface, along with heavy
quasiparticles whose dispersion never covers the entire Fermi
surface \cite{sc7,sc9,ragt,tamai,sigr,berr,puet}. Based on Eq.
\eqref{tem}, the temperature dependence of the effective mass
$M^*(T)$ of the FC quasiparticles is given by
\begin{equation} M^*(T)\sim
\frac{\eta p_F^2}{4T}, \label{M*}\end{equation} where $\eta=\delta
p/p_F$ is determined by the characteristic size $\delta p$ of the
momentum domain occupied by the FC and $p_F$ is the Fermi momentum
\cite{shagrep,shag,mig100}. From this relation it follows that the
effective mass of FC quasiparticles diverges at low temperatures,
while their group velocity, and hence their current, vanishes.
Therefore the main contribution to the resistivity is provided by
normal quasiparticles outside the FC having non-divergent effective
mass $M^*_{L}$ and finite group velocity at $T\to 0$.  Nonetheless,
it will be seen that FC quasiparticles still play a key role in
determining the behavior of the irregular residual resistivity in
the range $B_{c1}<B<B_{c2}$.

The resistivity has the conventional dependence \cite{pines}
\begin{equation}\label{rhoo}
\rho(T)\propto M^*_{L}\gamma
\end{equation}
on the effective mass $M^*_L$ and damping $\gamma$ of the normal
quasiparticles. Based on the relation \eqref{M*}, the behavior of
$\gamma$ is obtained in closed form in the present context as
\begin{equation}\label{LT}
\gamma\sim \eta(\gamma_0+T),
\end{equation}
where $\gamma_0$ is a constant \cite{prb,jetpl}. It is seen from
Eq. \eqref{LT} that the coefficient $A$ on the right hand side of
Eq. \eqref{res} is partly formed by FC. We call this contribution
to $A$, coming from FC, $A_{FC}$.

We consider now additional contribution $A_{TS}$ to the the
$T$-linear resistivity, formed by the zero sound generated by the
presence of FC \cite{DP,kpla}. The system with FC possesses its own
set of zero-sound modes. The mode of interest for our analysis
contributes to the $T$-linear dependence of the resistivity, as the
conventional sound mode does in the case of normal metals
\cite{kpla}. The mode is that of transverse zero sound with its
$T$-dependent sound velocity $c_t\simeq\sqrt{T/M_{\rm vHs}}$ and
the Debye temperature \cite{DP}
\begin{equation} T_D\simeq c_tk_{max}\simeq \beta\sqrt{TT_F}.\label{td}
\end{equation}
Here, $\beta$ is a factor, $M_{\rm vHs}$ is the effective mass of
electron formed by vHs, $T_F$ is the Fermi temperature, while $M^*$
on the left hand side of Eq. \eqref{M*} is the effective mass
formed finally by some interaction, e.g. the Coulomb interaction,
generating flat bands \cite{yudin,merge}. The characteristic wave
number $k_{max}$ of the soft transverse zero-sound mode is
estimated as $k_{max}\sim p_F$ since we assume that the main
contribution forming the flat band comes from vHs. We note that the
numerical factor $\beta$ cannot be established and is considered as
a fitting parameter, correspondingly, making $T_D$ given by Eq.
\eqref{td} uncertain. Estimating $T_F\sim 10$ K and taking
$\beta\sim 0.3$, and observing that the quasi-classical regime
takes place at $T>T_D\simeq \beta\sqrt{TT_F}$, we obtain that
$T_D\sim 1$ K and expect that strongly correlated Fermi systems can
exhibit a quasi-classical behavior with the low-temperature
coefficient $A$, entering Eq. \eqref{res}, $A=A_{FC}+A_{TS}$
\cite{DP,kpla,uni_r}.

Thus, in HF metals with their few bands crossing Fermi level and
populated by LFL quasiparticles and by HF quasiparticles, the
transverse zero sound make the resistivity possess the $T$-linear
dependence at the quantum criticality as the normal sound (or
phonons) do in the case of ordinary metals, while FC adds the
quantum contribution $A_{FC}$ to the coefficient $A$
\cite{kpla,uni_r}. The observed contributions lead to the lifetime
$\tau_q$, formed by normal, FC quasiparticles, and the transverse
zero sound,
\begin{equation}\label{LT1}
\hbar/\tau_q\simeq a_1+a_2T,
\end{equation}
where $\hbar$ is Planck's constant, $a_1$ and $a_2$ are
$T$-independent parameters. Relations \eqref{LT} and \eqref{LT1} is
in excellent agreement with recent experimental observations
\cite{tomph}. In playing its key role, the FC makes all
quasiparticles possess the same unique width $\gamma$ and lifetime
$\tau_q$. As we shall see, the $T$-independent width $\gamma_0$
forms the irregular residual resistivity $\rho_0^c$.

Using relations \eqref{res}, \eqref{rhoo} and \eqref{LT} together
with the standard treatment of vertex corrections \cite{trio}, we
are led to conclude that the resistivity of $\rm Sr_3Ru_2O_7$
should behave as
\begin{equation}
\rho\sim \rho_{\rm res}+\Delta\rho(B)+\rho_0^c+AT \label{rho}
\end{equation}
in the thermodynamic regime in question. The term ``residual
resistivity'' ordinarily refers to impurity scattering. In the
present case, as seen from Eqs. \eqref{LT} and \eqref{rho}, the
irregular residual resistivity $\rho_{0}^c$ is instead determined
by the onset of a flat band, and has no relation to scattering of
quasiparticles by impurities. Since the FC and the flat band
manifest themselves on in the region $B_{c1}<B<B_{c2}$, it is
natural also to conclude that the QCPs indicated in Fig.~\ref{fig2}
are responsible for the jumps in the irregular residual resistivity
$\rho_0^c$. According to the relations \eqref{LT} and \eqref{rho},
the resistivity $\rho$ is a linear function of $T$
\cite{prb,jetpl}. This feature of the flat-band scenario is in
accordance with the relevant measurements on $\rm Sr_3Ru_2O_7$
\cite{pnas}. Moreover, experimental observations and their
theoretical explanation show that the same physics describes the
$T$-linear dependence of the resistivity of conventional metals and
both HF metals and $\rm Sr_3Ru_2O_7$, with the quasi-classical
behavior formed by the zero-sound mode at their quantum criticality
\cite{bruin,uni_r}.

As it was discussed above, heavy quasiparticles shaped by the flat
band do not contribute directly to the transport properties.
Defining the lifetime $\tau_q$ instead, these specify the transport
of the system. As a result, the magnetoresistivity jumps and its
variation through the peak are defined by the variation of the
irregular residual resistivity $\rho_{0}^c$. Indeed, increasing the
temperature broadens and increases the resistivity in accordance
with Eq. \eqref{rho}, but the minimal values of the jumps and of
the peak exhibit the spectacular independence of temperature, as it
is seen from Fig. \ref{fig1} {\bf B}. A salient experimental
feature supporting this conclusion is the occurrence of two jumps
in the resistivity: first an upward jump at $B_{c1}$, where the FC
is built up, followed by a downward jump at $B_{c2}$, where the FC
is destroyed.  Thus, the scenario developed here reveals the
genesis of the two steep sidewalls observed in the irregular
residual resistivity $\rho_0^c$, in effects arising from the
formation of a flat band at QCPs. One could attribute $\rho_0^c$ to
the influence of the magnetic field $B$, considering $\rho_0^c$ as
a magnetoresistivity. Such a definition would obscure the physical
mechanism responsible for forming $\rho_0^c$. Indeed, it is the
flat band that forms the irregular residual resistivity, while the
magnetic field represents an auxiliary parameter that tunes the
system to the flat band.

\section{Entropy}

\begin{figure}[!ht]
\begin{center}
\includegraphics [width=0.47\textwidth]{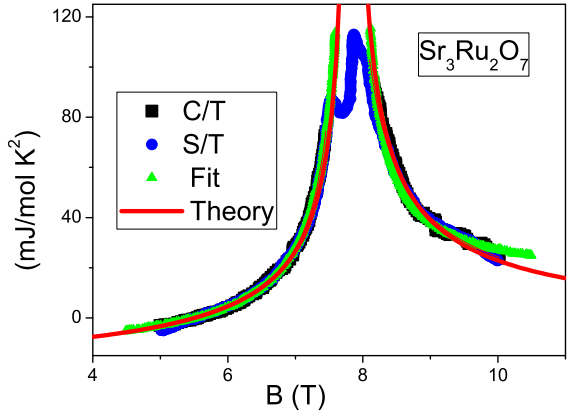}
\end{center}
\caption{(Color online). Magnetic-field dependence of both the
specific heat $C/T$ (depicted by the squares) and the entropy $S/T$
(circles) as obtained in measurements on $\rm Sr_3Ru_2O_7$ and
their divergent behavior as fitted by the function
$B_c|B-B_c|^{\alpha}$ with $\alpha=-1$ (triangles) \cite{sc9}. The
theoretical fit of the current work is shown as the solid curve
given by Eq. \eqref{SB}.} \label{fig3}
\end{figure}

In the LFL state depicted in Fig.~\ref{fig2}, both the entropy $S$
and the specific heat $C$ of the electron liquid in $\rm
Sr_3Ru_2O_7$ behave in accordance with LFL theory $S/T=C/T\propto
M^*$ \cite{trio}, with the exception that the effective mass $M^*$
depends on magnetic field $B$ according to $M^*(B)\propto
|B-B_c|^{-2/3}$ \cite{shagrep,khodprb}. Here, $B_c$ is the field at
which the QCP occurs; in the present case of two QCPs, $B_c$ is
taken equal to $B_c=(B_{c2}+B_{c1})/2\simeq 7.9$ T, for simplicity.
The entropy is then given by
\begin{equation}\label{SB}
S(B)/T=C/T\simeq A_s+D_s|B-B_c|^{-2/3},
\end{equation}
where $A_s$ and $D_s$ are fitting parameters for the low-field and
high-field QCPs. The LFL behavior of $C/T$ and $S/T$ fitted by Eq.
\eqref{SB} are shown by the solid curve in Fig.~\ref{fig3}, in
comparison with experimental results \cite{sc9} for these
quantities represented by squares and circle symbols, respectively.
The triangles display the fit with exponent $\alpha=-1$\,
\cite{sc9} rather than $\alpha-2/3$.  The two fits (the solid curve
and the triangles) are seen to show similar behavior as functions
of magnetic field. In contrast to the exponent $\alpha=-1$ obtained
by the fitting of the experimental data over small variation of the
magnetic field \cite{sc9}, the validity of the exponent
$\alpha=-2/3$ is then confirmed by the good agreement with the
experimental data and internal consistency between the schematic
phase diagram in Fig.~\ref{fig2} and the data in Figs. \ref{fig1},
\ref{fig3} and \ref{fig4}. Figure \ref{fig3} shows that the entropy
increases strongly on both the low-field and high-field sides of
the ordered phase as the critical fields $B_{c1}$ and $B_{c2}$ are
approached. Thus, the theory of FC allows us to explain, for the
first time, the experimental data collected for $\rm Sr_3Ru_2O_7$
\cite{sc9,gegen} on the evolution of the entropy and the heat
capacity as the quantum critical point is approached. We note that
the entropy jumps across the first-order phase transitions visible
in Fig.~\ref{fig3} are in accord with the phase diagram sketched in
Fig.~\ref{fig2}.

\section{Scaling behavior}

\begin{figure} [! ht]
\begin{center}
\includegraphics [width=0.40\textwidth]{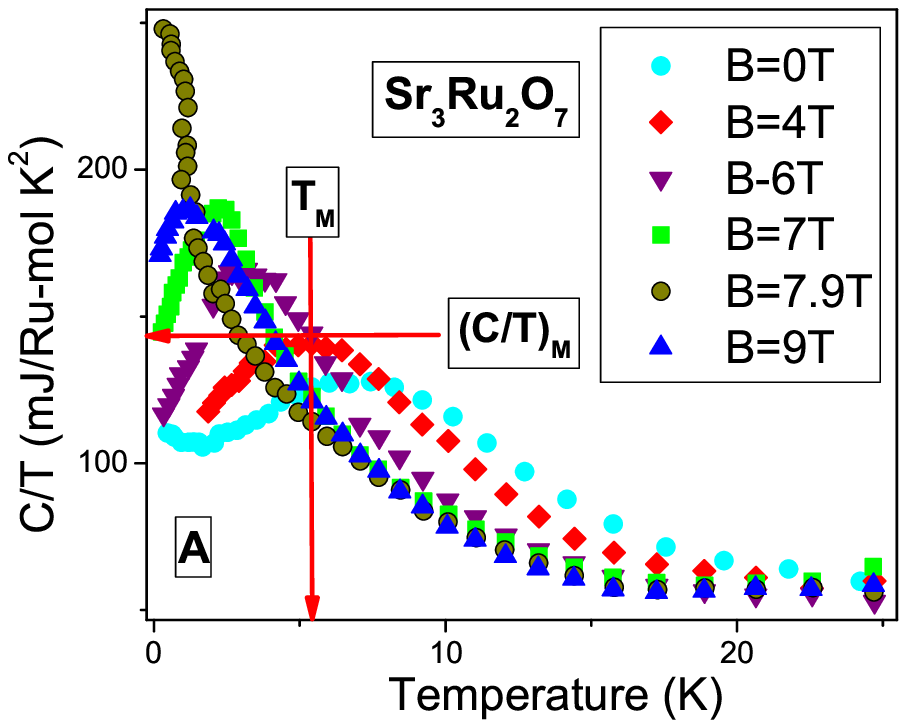}
\includegraphics [width=0.40\textwidth]{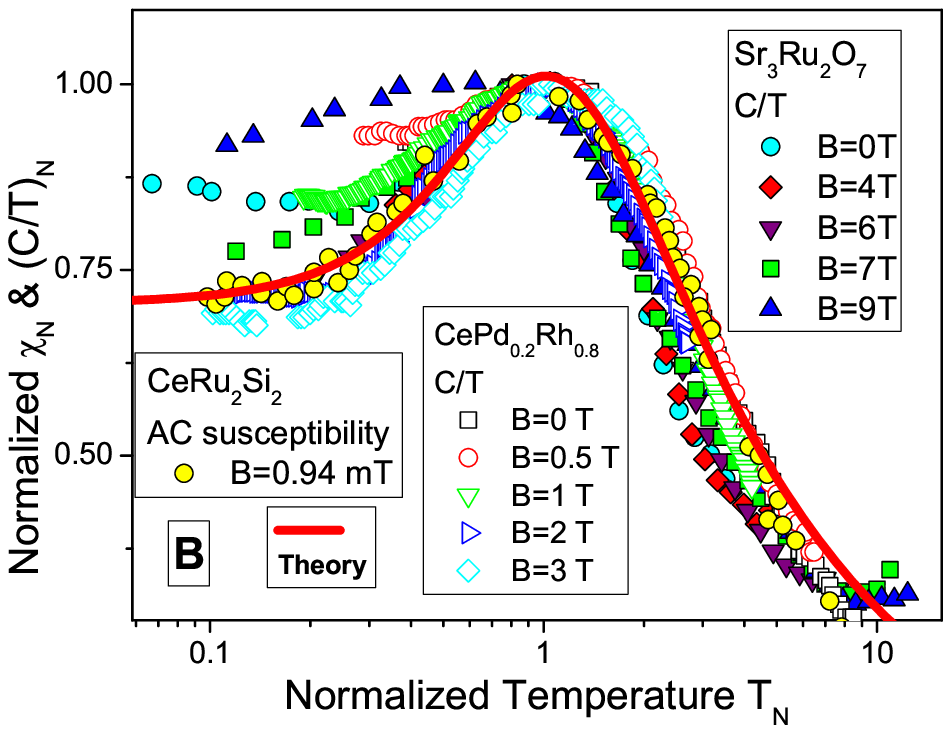}
\end{center}
\caption{(Color online). Panel {\bf A}. Temperature dependence of
the electronic specific heat for different magnetic field strengths
(after Rost {\it et al.} \cite{pnas}). The maximum at zero field
increases and shifts to lower temperatures as the magnetic field
approaches QCPs. At $B=7.9$ T no maximum occurs. A maximum
reappears on the high-field side of the transition, demonstrating a
symmetrical behavior with respect the critical region located
between the two QCPs and occupied by the ordered phase. The
illustrative values of $(C/T)_M$ and $T_M$ at $B=4$ T are shown by
the arrows. Panel {\bf B}. Universal scaling behavior of the
normalized specific heat $\chi_N=(C/T)_N\propto M^*$, extracted
from measurements on $\rm CeRu_2Si_2$ \cite{tak}, $\rm
CePd_{1-x}Rh_x$ with $x=0.80$ \cite{pik}, and $\rm Sr_3Ru_2O_7$
\cite{pnas}. All the measurements displayed in panels {\bf A} and
{\bf B} were performed under the application of magnetic fields as
shown in the insets. The solid curve represents our calculation of
the universal behavior. }\label{fig4}
\end{figure}

With the aim of revealing signatures of the hidden QCPs, we
conclude with an analysis of the thermodynamic properties of the
$C/T$ electronic specific heat measurements on $\rm
Sr_3Ru_2O_7$\,\, \cite{pnas}. As mentioned above, at $B=7.9$ T the
resistivity $\rho$ is precisely linear in $T$ over the range
$T_c\leq T<18$ K, with $C/T$ varying as $\ln T$ over the same range
\cite{pnas}. These are typical fingerprints of a flat band
generated by FC at QCPs \cite{prb,jetpl}. The experimentally
derived temperature dependence of $C/T\propto M^*$ on magnetic
field strength, shown in Fig.~\ref{fig4} {\bf A}, allows us to
uncover the universal scaling behavior of the effective mass $M^*$
characteristic of HF metals. As shown in this figure, the maximum
of $C/T\propto M^*$ sharpens and shifts to lower temperatures as
the field $B$ approaches 7.9 T, where the maximum disappears. In
contrast to HF metals, $C/T$ exhibits a symmetry with respect to
the area implicated by the ordering (nematic) transition: the
maximum appears upon approach to the QCPs and reappears on the
high-field side of this transition region. This behavior of the
maximum is in accord with the phase diagram of Fig.~\ref{fig2},
since the width $W(B)$ increases linearly with $T$ and the maximum
located in the transition region shifts toward zero temperature,
while the effective mass $M^*(B)$ given by Eq. \eqref{SB} diverges
as $B$ approaches the critical field.  To expose the scaling
behavior, we normalize the measured $C/T$ values to
$(C/T)_N=(C/T)/(C/T)_M$ and the corresponding temperatures $T$ to
$T_N$, $T_N=T/T_M$, by dividing by their values $T_M$ and $(C/T)_M$
at the maxima \cite{shagrep}. The elucidative values of $T_M$ and
$(C/T)_M$ at $B=4$ T are depicted by the arrows in panel {\bf A} of
Fig. \ref{fig4}. The spin $AC$ susceptibility data $\chi(T)\propto
M^*$ are normalized in the same way. At FQCPT, all the normalized
$\chi_N$ and $(C/T)_N$ curves are to merge into a single one,
$\chi_N=(C/T)_N=M^*_N(T_N)$, where $M^*_N$ is the normalized
effective mass represented by a universal function, being a
solution of the Landau equation \cite{shagrep}. This solution
$M^*_N(T_N)$ can be well approximated by a simple universal
interpolating function. The interpolation occurs between the LFL
and NFL regimes and represents the universal scaling behavior of
$M^*_N$ \, \cite{shagrep}
\begin{equation}M^*_N(T_N)\approx c_0\frac{1+c_1T_N^2}{1+c_2T_N^{8/3}}.
\label{UN}
\end{equation}
Here, $c_0=(1+c_2)/(1+c_1)$, $c_1$, and $c_2$ are fitting
parameters. Figure \ref{fig4} {\bf B} reports the behavior of the
normalized $\chi_N$ and $(C/T)_N$ thus extracted from measurements
on $\rm CeRu_2Si_2$\,\, \cite{tak}, $\rm CePd_{0.8}Rh_{0.8}$\,\,
\cite{pik} and $\rm Sr_3Ru_2O_7$\,\, \cite{pnas}. The solid curve
shows the result of our calculation of the scaling behavior that
can be well fit by Eq. \eqref{UN}. The HF metals and $\rm
Sr_3Ru_2O_7$ are seen to exhibit the same scaling behavior, which
can be understood within the framework of fermion condensation or
flat-band theory \cite{shagrep,shag,mig100}.\\

\section{Summary}

In summary, we have unveiled a challenging connection between $\rm
Sr_3Ru_2O_7$ and heavy-fermion metals by establishing universal
physics that straddles across the corresponding microscopic
details. Our construction of the $T-B$ phase diagram of $\rm
Sr_3Ru_2O_7$ has permitted us to explain main features of the
experimental one, and unambiguously implies an interpretation of
its extraordinary low-temperature thermodynamic in terms of fermion
condensation quantum phase transition leading to the formation of a
flat band at the restricted range of magnetic fields $B_{c1}\leq B
\leq B_{c2}$. We have demonstrated that the obtained agreement with
the experimental phase diagram is robust and does not depend on the
nature of the ordered phase, for our analysis is based on the
thermodynamic consideration. We have shown that it is the flat band
that generates both the entropy peak and the resistivity jumps, as
the critical fields $B_{c1}$ and $B_{c2}$ are approached. We have
also detected the scaling behavior of the thermodynamic functions
of $\rm Sr_3Ru_2O_7$ coinciding with that of heavy-fermion metals.

In the future, it would be interesting to analyze the physics of
the ordered (nematic) phase. Our preliminary results show that the
fermion condensation state breaks the discrete square lattice
rotational symmetry and generates a large magnetoresistive
anisotropy, as the magnetic field $B$ is rotated away from the $c$
axis toward the $ab$ plane.

\section{Acknowledgement}

Stimulating discussions with P. Gegenwart are gratefully
acknowledged. This work was supported by U.S. DOE, Division of
Chemical Sciences, Office of Basic Energy Sciences, Office of
Energy Research and AFOSR, and the McDonnell Center for the Space
Sciences.

\end{document}